\documentclass[10pt, reprint,aps,amsmath,amssymb,twocolumn,superscriptaddress]{revtex4-2}

\usepackage{graphicx}
\usepackage{bm}
\usepackage{color}
\usepackage{MnSymbol}
\usepackage{enumerate}
\usepackage{bbold,soul}
\usepackage{lipsum}
\usepackage{array}
\usepackage{hyperref}
\usepackage{textgreek}
\usepackage[dvipsnames]{xcolor}
\usepackage{tabularx}
\usepackage{graphicx}
\usepackage{cleveref}
\newcommand{\beq}{\begin{equation}}
\newcommand{\eeq}{\end{equation}}
\newcommand{\kd}{\kappa}%{\kappa\subd}
\newcommand{\kdvec}{\underline{\kappa}}%{\kappa\subd}
\newcommand{\ksvec}{\underline{K}}
\newcommand{\ks}{K}%{\kappa\subs}
\newcommand{\gu}{\gamma_\uparrow}%{\rm u}}
\newcommand{\gd}{\gamma_\downarrow}%{\rm d}

\renewcommand{\ul}[1]{\underline{{#1}}}

\newcommand{\ket}[1]{|{#1}\rangle}

\newcommand{\coh}{\mathcal{C}}
\newcommand{\cohc}{\coh^{\rm c}}
\newcommand{\cohnc}{\coh^{\rm nc}}

\newcommand{\phis}{\Phi}%{\phi\subs}
\newcommand{\phid}{\phi}%{\phi\subd}
\newcommand{\dd}{{\rm d}}

\newcommand{\ddt}{\tau} %\ddt t

\newcommand{\unxd}{\underline{X}}
\newcommand{\info}{Y}

\newcommand{\erf}[1]{Eq.~\eqref{#1}}

\definecolor{nblue}{rgb}{0.06,0.3,0.73}%229 11R, 61G, 145B
\definecolor{nblack}{rgb}{0,0,0}
\definecolor{nred}{rgb}{0.9,0.1,0.1}
\definecolor{nmagenta}{rgb}{0.7,0.0,0.3}
\definecolor{neditcolor}{rgb}{0.3,0.3,0.9}

\newcommand{\blk}{\color{nblack}}

\begin{document} 
\title{Multi-level Random-Telegraph Noise Mitigation using a Single Spectator Qubit}% Force line breaks with \\
% \thanks{A footnote to the article title}%

\author{Yanan Liu}
\email{yanan.liu@griffith.edu.au}
\affiliation{School of Engineering, University of Newcastle, Callaghan NSW 2308, Australia, \texorpdfstring{\\}{} Centre for Quantum Dynamics, Griffith University, Yuggera Country, Brisbane, Queensland 4111, Australia}

\author{Hongting Song}
\email{shtfc@163.com}
% \affiliation{Centre for Quantum Computation and Communication Technology (Australian Research Council), \texorpdfstring{\\}{} Centre for Quantum Dynamics, Griffith University, Yuggera Country, Brisbane, Queensland 4111, Australia}
\affiliation{Qian Xuesen Laboratory of Space Technology, China Academy of Space Technology,  Beijing 100094, China}%

\author{Areeya Chantasri}%
 \email{areeya.chn@mahidol.ac.th}
\affiliation{Optical and Quantum Physics Laboratory, Department of Physics, Faculty of Science, Mahidol University, Bangkok, 10400, Thailand}%
\affiliation{Centre for Quantum Computation and Communication Technology (Australian Research Council), \texorpdfstring{\\}{} Centre for Quantum Dynamics, Griffith University, Yuggera Country, Brisbane, Queensland 4111, Australia}

\author{Howard M. Wiseman}
\email{h.wiseman@griffith.edu.au}
\affiliation{Centre for Quantum Computation and Communication Technology (Australian Research Council), \texorpdfstring{\\}{} Centre for Quantum Dynamics, Griffith University, Yuggera Country, Brisbane, Queensland 4111, Australia}

\begin{abstract}
Preserving quantum coherence in the presence of environmental noise is one of the principal challenges for quantum technologies. Noise mitigation using spectator qubits (SQs) has recently emerged as a promising approach, enabling indirect probing of the noise without disturbing the data qubit (DQ). However, existing analyses that probe ultimate performance have been restricted to two-state random telegraph process noise, which does not capture more complex noise processes that may arise in the environment. Therefore, we here develop a SQ-based noise mitigation for DQs subject to general multi-level fluctuator noise. We first derive the coherence dynamics of the DQ under such noise, then develop a mitigation scheme in which information about the noise is inferred from sequential SQ measurements and used for phase correction. A memory-efficient heuristic adaptive protocol is proposed to dynamically select the SQ measurement time and angle based on the current noise estimate. Numerical simulations demonstrate that the proposed strategy significantly suppresses decoherence under multi-level noise, achieving performance comparable to that in the two-level case despite the increased complexity of the noise process. 
\end{abstract}

\maketitle

%\tableofcontents
\section{Introduction}
%\ach{The first 1.5 paragraphs of intro are too similar to that of the previous paper. Maybe we should think of something different? Like going from reviewing our previous work state-fluctator noise (non-Markovian, time-dependent noise, Bayesian and adaptive) and say that it's challenging to generalize it to arbitrary multi-state noise? We present the formalism that can be applied for any arbitrary noise values. Then some physics motivation as Yanan already wrote in the later part of 2nd paragraph.} \yl{I agree} 
Quantum technologies rely on the ability to preserve quantum coherence in the presence of environmental disturbances. However, physical qubits are inevitably affected by noisy environments, leading to decoherence and loss of quantum information. A wide range of noise mitigation techniques have been proposed, including dynamical decoupling (DD) \cite{viola1999dynamical,viola2003robust,biercuk2011dynamical,ng2011combining,souza2011robust,medford2012scaling,paz2013optimally,zhang2014protected} and quantum error correction (QEC)\cite{shor1995scheme,steane1996error,terhal2015quantum}. In this work, we are interested in mitigating time-dependent noise  using spectator qubits (SQs), where the SQs are sequentially probed over time to obtain information about the underlying noise process, which can be used to correct noise error and mitigate decoherence on the nearby data qubits (DQs)~\cite{song2023optimized,tonekaboni2023greedy,liu2026imperfection}.

For the DQ-SQ noise mitigation technique, both DQs and SQs are assumed to experience the same noise, but with different strengths. No direct measurements on the DQs are required, allowing the DQs to remain isolated as much as possible from the external environment. The SQs are instead assumed to be very sensitive to the noise and can be measured as often as required.  The SQ paradigm has recently been extended to a spectator-mode framework for mitigating spatially correlated noise~\cite{lingenfelter2023surpassing}. The DQ-SQ scheme has been experimentally demonstrated~\cite{singh2023mid} with two atomic species, where an array of Cesium atoms acting as SQs was used to correct phase errors on Rubidium DQs.

Previous treatments have proposed SQs algorithms that push towards the ultimate limits for this paradigm~\cite{song2023optimized,tonekaboni2023greedy,liu2026imperfection}, 
%\hmw{Are we really the first? Is the last one really RTP noise? Also it should be capital-H Heisenberg. Check all references for this.}\yl{the original means first?}\hmw{original. adjective.1 present or existing from the beginning; first or earliest: the original owner of the house | the plasterwork is probably original.2 created personally by a particular artist, writer, musician, etc.; not a copy: original Rembrandts.3 not dependent on other people's ideas; inventive or novel: a subtle and original thinker.} 
by focussing on the simplest case of a two-state random telegraph process (RTP) noise. Here the RTP, which appears multiplicatively in the qubit Hamiltonians, switches stochastically between two values. This type of noise has been widely observed, such as noise arise from a single charge fluctuator causing dephasing in charge qubits~\cite{hu2006charge,shalak2023modeling} and from defects in spin qubits and superconducting qubit platforms~\cite{zwanenburg2013silicon}. 
While the RTP model provides useful intuition and analytical tractability, experimental studies indicate that realistic noise environments often involve contributions of multiple RTPs, leading to a general multiple state fluctuator model~\cite{paladino20141}. In solid-state systems, ensembles of charge traps or defects, such as interface states, oxide traps, or impurities in the substrate, can simultaneously couple to the qubit, each with distinct switching rates and coupling strengths \cite{galperin2004low}. The combined effect of these fluctuators leads to complex, multi-level or effectively continuous noise processes, often associated with non-Gaussian statistics and $1/f$-type spectra~\cite{paladino20141}. 
%\hmw{okay to move here to slim down the presentation?} 
These observations suggest that modeling the environment as a single two-level fluctuator is insufficient to capture the full noise characteristics in realistic systems. Consequently, designing noise-mitigation strategies that can operate effectively under multi-level fluctuator noise processes is an important and largely open problem.

In this work, we investigate noise mitigation for DQs subject to general multi-level fluctuator noise. The noise process is modeled as a stochastic process taking values from $L$ levels, for arbitrary $L$, and undergoing random transitions between these states. With the two-state RTP noise assumption, $L=2$, the generalized framework of the Bayesian map-based method in~\cite{song2023optimized,tonekaboni2023greedy,liu2026imperfection} allow us to compute the DQ's coherence and likelihood of unknown noise values conveniently with $2\times 2$ matrices and $2\times 1$ vectors. In an appropriate regime, this yielded an analytical expression for decoherence suppression by our algorithm, which we conjectured to be optimal.  
Here, we extend the map-based formalism to arbitrary $L > 2$ noise levels, with $L \times L$ matrices and $L \times 1$ vectors. Many of the same considerations apply, but the higher complexity makes fully analytical results infeasible. 

We first derive the coherence dynamics of the DQ under such noise and analyze the structure of the associated fluctuation process. Building on this framework, we explore the use of a SQ to extract information about the underlying noise process and mitigate decoherence via adaptive measurements. 
%\ach{we have to come up with jusitification why we can only present a heuristic algorithm? More to discuss with Howard. No optimal solutions?} 
In particular, we propose a heuristic measurement strategy that dynamically selects the SQ measurement time and angle based on the current noise estimate. Numerical simulations demonstrate that the proposed strategy can suppress the decoherence rate of the DQ under multi-level noise to a level comparable to that achieved in the two-state RTP case.

The remainder of this paper is organized as follows. Section \ref{sec:mathmodel} presents all the necessary mathematical formulation, which include the general multi-level noise model in subsection \ref{sec:multilevelnoisemodel}, the DQ's coherence calculation without control in subsection \ref{sec:nocontrol}, and the coherence calculation with the SQ in subsection \ref{sec-dqsq}. In Section \ref{sec:heuristic_adaptive_protocol}, we present the heuristic adaptive strategy and derive an analytical upper bound for the DQ's decoherence under this strategy. 
Section \ref{sec:numericsim} presents the numerical simulation results on 3-level noise, and Section \ref{sec:conclusion} concludes the paper.

\section{DQ-SQ formalism for general multi-level noise}\label{sec:mathmodel}
In this section, we give the necessary mathematical description of the DQ-SQ formalism, here generalized for arbitrary multi-level noise models. We start from the definition of multi-level noise model~\ref{sec:multilevelnoisemodel}, followed by the DQ-SQ Hamiltonians and their evolutions~\ref{sec:dqsqham}. We then show how to compute the no-control coherence of the DQ under this noise in \ref{sec:nocontrol}. We extend the calculation of the DQ's coherence to the case with SQs and measurement strategies in \ref{sec-dqsq}. Then in subsection \ref{sec:cohvector}, we summarize the coherence vector in the map-based formalism which will be used extensively in the heuristic adaptive protocol for noise mitigation.

\subsection{Multi-level random-telegraph noise}\label{sec:multilevelnoisemodel}

We generalize the two-state Random-Telegraph Process (RTP) noise considered in the previous paper~\cite{song2023optimized,tonekaboni2023greedy,liu2026imperfection} to a model of a multi-state fluctuator noise process $m_t$, which can take, without loss of generality, integer values from 1 to $L$ at any time, i.e., $m_t \in \{ 1, 2, \cdots, L \}$.  
%\hmw{There is no point defining $1$, $2$ etc. We should just have $l$ taking integer values, and we an say that in that earlier work we took $l$ to have values $\pm$, because that gave more elegant formulas.} 
Let us assume that the noise value $m_t$ switches between these $L$ levels, with transition rates denoted by $\gamma_{ij}$, for the transitions from levels $i$ to $j$, where $i,j \in \{1, 2,...,L\}$. Since these noise values are unknown, we instead describe a state of noise by a probability vector at any time $t$ as
\begin{equation}\label{eq:probofz}
    \underline{P}_t=\begin{bmatrix}
        \wp(m_t=1)\\
        \wp(m_t=2)\\
        \vdots\\
        \wp(m_t=L)
    \end{bmatrix},
\end{equation}
where $\wp(m_t=i)$ is the probability for $m_t$ to be at the noise level $i$ at time $t$. The evolution of the probability vector $\underline{P}_t$ is governed by the master equation:
\beq
\partial_t\underline{P}_t=J\underline{P}_t,\label{eq:mastereq}
\eeq
where $J$ is the $L \times L$ generator matrix describing all possible transitions among the $L$ levels: 
\begin{equation}\label{eq:Jmatrix}
    J =
\begin{bmatrix}
-\sum_{j\neq 1}\gamma_{1j} & \gamma_{21} & \cdots & \gamma_{L1}\\
\gamma_{12} & -\sum_{j\neq 2}\gamma_{2j} & \cdots & \gamma_{L2}\\
\vdots & \vdots & \ddots & \vdots\\
\gamma_{1L} & \gamma_{2L} & \cdots & -\sum_{j\neq L}\gamma_{Lj}.
\end{bmatrix}
\end{equation}
%\begin{equation}\label{eq:Jmatrixalt}
%J_{ij} =
%\begin{cases}
%\gamma_{ij}, & i \neq j, \\
%-\sum\limits_{k \neq i} \gamma_{ik}, & i = j.
%\end{cases}
%\end{equation}
% \begin{equation}\label{eq:Jmatrix}
%     J=\left[\begin{smallmatrix}
%         \gamma_{1j}-\gamma_{12}-\gamma_{13}-\cdots -\gamma _{1L}&\gamma_{21}&\cdots&\gamma_{L1}\\
%         \gamma_{12}&-\gamma_{21}-\gamma_{23}-\cdots-\gamma_{2L}&\cdots&\gamma_{L2}\\
%         \vdots& \vdots& \vdots&\vdots\\
%         \gamma_{1L}&\gamma_{2L}&\cdots&-\gamma_{L1}-\gamma_{L2}-\cdots-\gamma_{L(L-1)}
%     \end{smallmatrix}\right]
% \end{equation}
The master equation Eq.~\eqref{eq:mastereq} has a closed-form solution, for a finite-size $L$, between any time $t^\prime$ and $t$:
\begin{equation}\label{eq:Ptsolution}
    \underline{P}_t=e^{J(t-t^{\prime})}\underline{P}_{t^{\prime}}.
\end{equation}
We can also compute a steady state $\ul{P}_{\rm ss}$, such that $\partial_t\ul{P}_t = 0$.
%Compared with the two-level noise case, the above solution 
For a large $L$, analytical solutions for steady states can be obtained only for special cases, e.g., when the transition rates are symmetric, $\gamma_{ij} = \gamma_{ji}$. In most cases, we compute the steady states  numerically.
%\hmw{why the short form and the quote marks? Also, do we assume that in this paper? If so, is it justified? If not, why bother?} \yl{no, we donot assume this in the paper but the following steady state uses this assumption and the steady state is used in this paper}
%We have presented a detailed calculation for the steady state in Appendix \ref{sec_app_Jmatrix}. %This means that if we start from the uniformly distributed probability $P_0$, without any control, the long-time steady state of the noise process will be in the same uniformly distributed.

\subsection{DQ-SQ Hamiltonians and state evolutions}\label{sec:dqsqham}

Following the proposed model for noise mitigation~\cite{song2023optimized,tonekaboni2023greedy,liu2026imperfection}, we assume that both the DQ and SQ experience the same time-dependent noise $m_t$, but the SQ can be easily accessed and measured by the experimenter to extract the noise information, which then be used to correct any error occurred in the DQ. We consider a Hamiltonian of the DQ and SQ of the form
\begin{equation}\label{eq:totalHamiltonian}
    \hat{H}_{\rm tot}(t) = \frac{\tilde\kd(t)}{2} \hat{\sigma}_z^{\rm d} + \frac{\tilde\ks(t)}{2} \hat{\sigma}_z^{\rm s} + \hat H_{\rm ctrl}^{\rm d}\delta(t-T),
\end{equation}
where the first and the second terms on the right describe the noise effects on the DQ's and SQ's with the corresponding $z$-Pauli operators, $\hat\sigma_z^{\rm d}$ and $\hat\sigma_z^{\rm s}$. Here $\tilde\kd(t)$ and $\tilde\ks(t)$ are the qubits' frequency shifts, for the DQ and SQ respectively, that depend on the noise values $m_t$ at time $t$.
%For convenience, we assume that the sensitivities take $L$ different values $\kd(t) \in \{ \kappa_1, \kappa_2, \cdots, \kappa_L\}$ and $\ks(t) \in \{ K_1, K_2, \cdots, K_L\}$ corresponding to the noise value $z(t)$ at that time. That is,
%\begin{equation}\label{eq:kappsl}
%    \kd_j=\kd(z(t)=l_j)l_j, j=1,2,\cdots,L.
%\end{equation}
%\begin{align}\label{eq:kskd}
%    \kd(t) =\, \kd_j \text{ and } \ks(t) =\, \ks_j, \text{ when } z(t) = l_j
%\end{align}
%Here, $\kd(z(t))$ is the product of sensitivities and the corresponding noise values for the DQ, and $\ks(z(t))$ is that of the SQ.
For the last term in Eq.~\eqref{eq:totalHamiltonian}, the Hamiltonian describes the error-correction control that can be applied on the DQ at a final time $T$.
%\begin{equation}\label{eq:totalHamiltonian}
%    \hat{H} = \frac{\kd(z(t))}{2} \hat{\sigma}_z^{\rm d} \, z(t).
%\end{equation}
%Here, $\kd(z(t))$ is the sensitivity function that changes with the value of $z(t)$.
Given the Hamiltonian in Eq.~\eqref{eq:totalHamiltonian}, the RTP $m_t$ affects only the phases of the DQ and SQ.
%and $\kd_j$ in \eqref{eq:kappsl} denotes the phase changed by the noise level $l_j$. 
Therefore, to calculate the effect of this noise, 
%without loss of generality, 
we can consider the equatorial states of the DQ,
\begin{equation}\label{eq:dqstate}
    \ket{\phi}^{\rm d} := \frac{1}{\sqrt{2}} \left(\ket{+1}_z^{\rm d} + e^{i \phi} \ket{-1}_z^{\rm d} \right),
\end{equation}
where $\ket{\pm 1}_z^{\rm d}$ are the eigenstates of the $z$-Pauli matrices of the DQ. Similarly for the SQ, to obtain maximum sensitivity we want to prepare an equatorial state, which then evolves to 
\begin{equation}\label{eq:sqstate}
\ket{\phis}^{\rm s} := \frac{1}{\sqrt{2}} \left(\ket{+1}_z^{\rm s} + e^{i \phis} \ket{-1}_z^{\rm s} \right).
\end{equation}
%where  $\ket{\pm 1}_z^{\rm d}$ are the eigenstates of the $z$-Pauli matrices of the SQ. 

Let us set both DQ's and SQ's states to be the zero-phase states, $\ket {\phid=0}^{\rm d}$ and $\ket {\phis=0}^{\rm s}$, at the initial time. The qubits' frequency shifts $\tilde\kd(t)$ and $\tilde\ks(t)$ 
can be written, without loss of generality, as 
%will have values depending on which states of noise at time $t$ and how much the qubits are sensitive to the noise, i.e.,
\begin{subequations}\label{eq-effectfreq}
\begin{align}
    \tilde\kd(t) &= \sum_{j=1}^L \delta_{m_t,j} \kd_j \,,\\
    \tilde\ks(t) &= \sum_{j=1}^L \delta_{m_t,j} \ks_j\,,
\end{align}
\end{subequations}
where we have used the Kronecker-$\delta$ function $\delta_{m_t,j}$ and defined the noise sensitivities $\kd_j$ and $\ks_j$ for all noise states $j \in \{ 1,2,..., L\}$. We take the list of $\kd_j$s to be non-degenerate. If the DQ had the same response to two RTP levels then they might as well be treated as the same level. We also take the list of $\ks_j$s to be non-degenerate. If this were not the case, there would be a more serious problem in that the SQ would be unable to distinguish between levels of the noise that do affect the DQ differently. We will return to this issue in the Conclusion.

Therefore, under the Hamiltonian evolution (prior to $T$), the states will evolve from $t=0$ to time $t$ according to: 
\begin{subequations}
\begin{align}\label{eq:dataphase}
\ket {\phid(\ul{X})}^{\rm d} = &\,\exp\left(- \tfrac{\underline{i}}{2}\hat \sigma_z^{\rm d} \! \int_0^{t} \! \dd s\,  \tilde\kd(s) \right)\ket {\phid=0}^{\rm d}, \\
%=&\,\ket {\underline{\kd}^{\top} \underline{X}}^{\rm d},\\
|\Phi(\ul{X})\rangle^{\rm s} = &\,\exp\left(- \tfrac{\underline{i}}{2}\hat \sigma_z^{\rm s} \! \int_0^{t} \! \dd s\,  \tilde\ks(s) \right)\ket {\phis=0}^{\rm s}.
% =&\, |\ul{\ks}^{\top} \ul{X} \rangle^{\rm s},
\end{align}
\end{subequations}
Here, by substituting in Eq.~\eqref{eq-effectfreq}, we 
have written the DQ's and SQ's phases as 
%\hmw{I suggest leaving out the 3rd term in both of the equations below as we define it later anyway.} \ach{The third terms? But they were simply from substituting (8a) and (8b).}\hmw{I know. They are not wrong. I'm saying they are superfluous. They make the equations that follow seem a bit pointless.}
\begin{subequations}\label{eq:phases}
\begin{align}
\phid(\ul{X}) =& \int_0^{t} \! \dd s\,  \tilde\kd(s) 
%= \sum_{j=1}^L \kd_j \!\!\int_0^{t} \! \dd s\,\delta_{m_s,j} 
\equiv \ul{\kd}^{\top} \ul{X},\\
\Phi(\ul{X}) =& \int_0^{t} \! \dd s\,  \tilde\ks(s) 
%= \sum_{j=1}^L \ks_j \!\!\int_0^{t} \! \dd s\,\delta_{m_s,j}
\equiv \ul{\ks}^{\top} \ul{X},
\end{align}
\end{subequations}
where we have defined noise sensitivity vectors
\begin{subequations}\label{eq:sensitivity}
\begin{align}
    \kdvec=&(\kappa_1,\kappa_2,\cdots,\kappa_L)^{\top},\\
    \ksvec=&(K_1,K_2,\cdots,K_L)^{\top},
\end{align}
\end{subequations}
and an accumulated noise-time vector 
\beq
\underline{X}=(X_1,X_2,\cdots,X_L)^{\top},
\eeq
such that
\beq\label{eq:Xvalue}
X_j(t)=\int_0^{t}\! \dd s\, \delta_{m_s,j}.
\eeq
That is, the value $X_j(t)$ represents the amount of time  that $m_s = j$ in the period $s \in (0,t]$.

It is worth commenting on the connection of the above formulation to that used in the previously considered~\cite{song2023optimized,tonekaboni2023greedy,liu2026imperfection} special case,  
%of the multi-level noise formalism presented here, 
the two-state RTP, which has only two values, $m_t \in \{1,2\}$. In that previous work, the notation was slightly different. For example, for the SQ, the authors used $\tilde\ks(t) = \ks\, z(t)$, where $z(t) \in \{ +1, -1\}$ and $\ks$ is a constant noise sensitivity. This means that $\tilde\ks(t)$ can take two values with opposite signs, i.e., $\ks_1 = \ks$ and $\ks_2 = -\ks$. Also, the noise switching rates were denoted $\gd = \gamma_{12}$ and $\gu=\gamma_{21}$. We note that the definition of $X_j$ in Eq.~\eqref{eq:Xvalue} is more subtle than the accumulated noise `$X$' in the previous work. Therefore, using the definition in Eq.~\eqref{eq:Xvalue}, we find that $X_1$ and $X_2$ in the current notation are the total time that $m_s$ stays in the level 1 and 2 during the time period $s\in (0,t]$, and the effective sensitivities in Eq.~\eqref{eq:sensitivity} are $\kd_1 = -\kd_2 = \kappa$ and $\ks_1 = -\ks_2 = K$ for the DQ and SQ, respectively.

We note that we have simplified the problem such that the only unknown variable is the vector of accumulated noise times $\ul{X}$, which results in the unknown phases $\phid(\ul{X})  = \ul{\kd}^{\top} \ul{X}$ and $\Phi(\ul{X})= \ul{\ks}^{\top} \ul{X}$. In the following, we will explain how the unknown phase of the DQ leads to its decoherence and how we can estimate the value of $\ul{X}$ by measuring the SQ's phase.

%\ach{move it somehwere} Since we assume $\ks(z(t))\gg \kd(z(t))$ for all $z(t)$, we have $\ks_j\gg \kd_j$. $\hat H_{\rm ctrl}^{\rm d}$ is the control Hamiltonian that is only applied at the final time, $T$.

\subsection{DQ's coherence with no control}\label{sec:nocontrol}

Since the noise variable $\ul{X}$ is unknown, the DQ's phase $\phid(\ul{X})  = \ul{\kd}^{\top} \ul{X}$ is a random variable. We can compute the DQ's coherence from an average over the phase factor:
\begin{equation}\label{eq:cohenocon}
    \mathcal{C}^{\rm nc}=\left|\langle e^{i\kdvec^{\top}\underline{X}}\rangle_{\underline{X}}\right|=\left|\int \wp(\underline{X})e^{i\kdvec^{\top}\underline{X}} \dd \ul{X}\right|,
\end{equation}
where the expectation value is over all possible values of the vector $\ul{X}$. That is $\wp(\ul{X})=\wp(X_1,X_2,\cdots,X_L)$ is a joint probability of all $X_j$ with the integral measure $\dd \ul{X} = \dd X_1 \dd X_2 \cdots \dd X_L$. The `nc' superscript denotes the case when there is no control Hamiltonian $\hat H_{\rm ctrl} = 0$ and simply letting the DQ experience the noise over time until the final time $t=T$. We apply the map-based formalism~\cite{song2023optimized,tonekaboni2023greedy,liu2026imperfection} as detailed in Appendix~\ref{sec_app_Hmatrix}, to evaluate the no-control coherence of the DQ at any time $t$ as
\begin{equation}\label{eq:cohenoconH}
    \mathcal{C}^{\rm nc}=\left| \underline{I}^{\top} {\bf H}(t,\ul{\kd})\ul{P}_0 \right|.
\end{equation}
Here $\underline{I}=(1, \cdots, 1)^{\top}$ is a vector of unit values of size $L$  and $\ul{P}_0$ is the probability vector Eq.~\eqref{eq:probofz} at time $t=0$. We can compute the matrix ${\bf H}(t, \ul{\kd})$ by discretizing the time to $t = M \Delta t$ taking $M \rightarrow \infty$, 
which gives the following form: 
%\hmw{Why do we need the pre- and post-multiplying matrices here? They are basically the identity in the limit $M\to\infty$.} \ach{This is exactly what we did with $L=2$ papers, the limit should be taken after the multiplication is done. Now I'm curious if we should write it here... it's too much detail}\hmw{But see my comment below.} \ach{maybe move to appendix? In the previous papers, they were all in appendices.}\hmw{Okay. Is there a differential equation that ${\bf H}$ obeys? We could put that here. And our numerical solution in the Appendix.} \ach{not a DE but a complicated summation over all L levels at all time dt steps. Not trivial. See Eq. (61) in Appendix just to have an idea.}\hmw{I see. Well let's just leave these defs here. But I still don't see why in (!6) we need the outside matrices. Since they are already there in (17).}\yl{no, (16) and (17) have different pre- and post-multiplying matrices}\hmw{oh. Okay.}
\begin{widetext}
\begin{equation}\label{eq:Hmatrixapp}
    {\bf H}(t, \ul{\kd})=\lim_{M\rightarrow\infty}\begin{bmatrix}
        e^{i\kd_1\Delta t/2}&0&\cdots& 0\\
        0&e^{i\kd_2\Delta t/2}&\cdots&0\\
        \vdots&\vdots&\ddots&\vdots\\
        0&0&\cdots&e^{i\kd_L\Delta t/2}
    \end{bmatrix}{\bf M}^M
    \begin{bmatrix}
        e^{-i\kd_1\Delta t/2}&0&\cdots& 0\\
        0&e^{-i\kd_2\Delta t/2}&\cdots&0\\
        \vdots&\vdots&\ddots&\vdots\\
        0&0&\cdots&e^{-i\kd_L\Delta t/2}
    \end{bmatrix},
\end{equation}
%\end{widetext}
%Here $\bf{M}$ is calculated as:
%\begin{widetext}
where the matrix ${\bf M}$ can be obtained from
\begin{equation}\label{eq:bfMapp}
    \bf{M}=\begin{bmatrix}
        e^{i\kd_1\Delta t/2}&0&\cdots&0\\
        0&e^{i\kd_2\Delta t/2}&\cdots&0\\
        \vdots&\vdots&\ddots&\vdots\\
        0&0&\cdots&e^{i\kd_L\Delta t/2}
    \end{bmatrix}\begin{bmatrix}
        T_{11}&T_{12}&\cdots& T_{1L}\\
        T_{21}&T_{22}&\cdots& T_{2L}\\
        \vdots&\vdots&\ddots&\vdots\\
        T_{L1}&T_{L2}&\cdots&T_{LL}
    \end{bmatrix}\begin{bmatrix}
        e^{i\kd_1\Delta t/2}&0&\cdots&0\\
        0&e^{i\kd_2\Delta t/2}&\cdots&0\\
        \vdots&\vdots&\ddots&\vdots\\
        0&0&\cdots&e^{i\kd_L\Delta t/2}
    \end{bmatrix},
\end{equation}    
\end{widetext}
and we have defined $T_{ij}=\wp(m=i|m^\prime=j) = [\exp(J \Delta t)]_{ij}$ as a transition matrix element following Eq.~\eqref{eq:Ptsolution}. We note that for the two-level RTP noise, $L = 2$, the analytical expression of $\bf{H}$ can be obtained through the eigenvalue decomposition. However, for $L>2$, the matrix equation becomes too complicated for analytical calculation. In this work, we thus focus on numerical calculation of matrix multiplication by taking $M$ to be large enough.

\subsection{DQ's coherence and control with SQ's measurements}\label{sec-dqsq}

As we calculated in Eq.~\eqref{eq:cohenocon} and Eq.~\eqref{eq:cohenoconH}, the coherence of the DQ will likely decay over time from the random multi-level noise values. However, if we can obtain information about the noise, through measuring the SQ, a proper correction action can be applied to the DQ when it is needed. We will measure the SQ, which is expected to be much more sensible to the noise than the DQ, to obtain the noise information to correct any phase error and reduce the decoherence on the DQ.

To acquire information about the noise, let us probe the SQ at different times throughout the process. We denote the measurement times as:
\begin{equation}\label{eq:probetimesdef}
    t_1,t_2,\cdots,t_n,\cdots,t_{N-1},t_{N}=T.
\end{equation}
We assume that the SQ is initialized at the zero-phase state, $|\Phi=0\rangle^s$, at the initial time $t_0=0$, and after every  measurement, such that the noise over the time intervals of the duration
\begin{equation}\label{eq:taun}
\tau_n:=t_n-t_{n-1}  ,
\end{equation}
can be independently probed. Following Eq.~\eqref{eq:dataphase}, for a measurement interval between $t_{n-1}$ and $t_n$, the SQ will pick up a phase as described by
\begin{align}
    |\Phi(t_n)\rangle^{\rm s} =|\ul{\ks}^{\top}\ul{x}_n\rangle^{\rm s}
\end{align}
where $\ul{x}_n=(x_{n}^{(1)},x_{n}^{(2)},\cdots,x_{n}^{(L)})^\top$ is a vector of accumulated noise time similar to Eq.~\eqref{eq:Xvalue} but only during the interval $(t_{n-1}, t_n]$, i.e.,
\begin{align}
    x_{n}^{(j)}=\int_{t_{n-1}}^{t_n} \dd s \, \delta_{m_s, j} 
\end{align}

We then measure the SQ state projectively at times $t_1, t_2, \cdots, t_N=T$, with the projector:
\begin{equation}\label{eq:measurementangle}
    \hat{\theta}_n=\mathbb{1}-|\theta_n\rangle^{\rm s}\langle\theta_n|,
\end{equation}
where $\mathbb{1}$ is the identity operator and $$|\theta_n\rangle^{\rm s} := 1/\sqrt{2}(|+1\rangle_z^{\rm s}+e^{i\theta_n}|-1\rangle_z^{\rm s}).$$ Using the Born's rule, we find that the probability of outcome $y_n \in \{ 0, 1 \}$ is given by
\begin{equation}
    \wp(y_n|\theta_n,\ul{x}_n)=y_n+(-1)^{y_n}\cos^2\left[ \frac{1}{2}(\theta_n-\ul{\ks}^{\top}\ul{x}_n) \right].
\end{equation}
Therefore, by measuring the SQ, one can try to estimate the unknown vector $\ul{x}_n$. Since there are in total $N$ SQ's measurements, we obtain a string of records $Y := \{ y_1,y_2,\cdots,y_N\}$, which will be used to estimate the string of $ \ul{x}_1, \ul{x}_2, ..., \ul{x}_N$ and eventually the total accumulated noise time, i.e.,
\begin{align}
    \ul{X} = \sum_{n = 1}^{N} \ul{x}_n,
\end{align}
which is a vector of size $L$, representing the accumulated noise time at the final time $t_N = T$.

With all the measurement results, we can estimate the correcting phase as $c(Y)$, which will maximize the coherence of the DQ:
\begin{align} \label{eq:oldcoh_ctrl}
    \coh_{c(Y)} :=& \Big|\big\langle e^{i [\ul{\kd}^{\top} \unxd - c(\info)]}\big\rangle_{\unxd,Y} \Big|,\notag\\
    =&\bigg| \sum_{Y} e^{-i c(Y)} \wp(Y)\int \!\! \dd \unxd \,  e^{i \ul{\kd}^{\top} \unxd}    \wp(\unxd|Y)  \bigg|, \notag\\
     \le &\, \sum_{Y}  \wp(Y)\left| \int \!\! \dd \unxd \,  e^{i \ul{\kd}^{\top} \unxd}   \wp(\unxd|Y) \right|,
\end{align}
where, in the last line, the inequality becomes an equality when 
\beq\label{eq:optctrl}
c(Y) =\arg  \int \!\! \dd \ul{X} \,  e^{i \ul{\kd}^{\top} \unxd}   \wp(Y) \wp(\unxd|Y) = \arg \big\langle e^{i \ul{\kd}^{\top} \unxd} \big\rangle_{\unxd|Y}.
\eeq
That is, the coherence is maximized if the phase correction $\hat H_{\rm ctrl} = -c(Y) \hat \sigma_z^{\rm d}/2$ is applied to the DQ at the final time. By substituting $c(Y) = \arg \big\langle e^{i \ul{\kd}^{\top} \unxd} \big\rangle_{\unxd|Y}$ in Eq.~\eqref{eq:oldcoh_ctrl}, we obtain the phase-corrected coherence: 
\begin{align} \label{eq:maxcoh}
    \cohc := & \,\sum_{\info}  \wp(\info)\left| \,  \big\langle e^{i \ul{\kd}^{\top} \unxd} \big\rangle_{\unxd|\info}\right|\\
    = &\,\sum_{Y}\left| \underline{I}^{\top} {\bf F}(\mu_N,y_N)\cdots {\bf F}(\mu_1,y_1)\underline{P}_0 \right|\label{eq:maxcoh2},
\end{align}
where, in the second line, we have applied the map-based formalism~\cite{song2023optimized,tonekaboni2023greedy,liu2026imperfection} where $\ul{P}_0$ is the probability vector Eq.~\eqref{eq:probofz} at time $t=0$ and the measurement map,
\begin{align} \label{eq:Fmatrix}
    {\bf F}(\mu_n,y_n)=\,\frac{1}{4}\Big[& 2{\bf H}(\ddt_n,\ul{\kd})+(-1)^{y_n}e^{-i\theta_n} {\bf H}(\ddt_n,\ul{\kd}+\ul{\ks})\nonumber \\
    &\, +(-1)^{y_n}e^{+i\theta_n} {\bf H}(\ddt_n,\ul{\kd}-\ul{\ks})\Big],
\end{align}   
describes effect of the measurement at time $t_n$ with a setting $\mu_n=\{\theta_n,\ddt_n\}$ representing the measurement angle $\theta_n$ and waiting time $\ddt_n$. 
The full calculation details of the map-based coherence in the previous work~\cite{song2023optimized,tonekaboni2023greedy,liu2026imperfection} can still be applied, with simple generalization to $\ul{\kd}$ and $\ul{\ks}$ vectors. Interestingly, for $L = 2$ case, we find that the current version of ${\bf H}$ (and also ${\bf F}$) is related to the original version in Refs~\cite{song2023optimized,tonekaboni2023greedy,liu2026imperfection}, denoted by $\tilde{\bf H}$ and $\tilde{\bf F}$ for now, by these simple relations
\begin{align}\label{eq:oldnewH}
    {\bf H}(\ddt_n, \ul{\kd}) = &\, e^{i\alpha'}  \tilde{\bf H}(\ddt_n, (\kd_1 - \kd_2)/2),\\
    {\bf H}(\ddt_n, \ul{\kd} \pm \ul{\ks}) =& \, e^{i\alpha''_\pm} \tilde{\bf H}(\ddt_n, (\kd_1 - \kd_2)/2 \pm (\ks_1 - \ks_2)/2),\nonumber \\
    {\bf F}(\{\theta_n, \tau_n\}, y_n) =&\, e^{i\alpha'} \tilde{\bf F}(\{\theta_n - (\ks_1 + \ks_2)\tau_n/2, \tau_n\}, y_n).\nonumber
\end{align}
The global phases, $\alpha' = (\kd_1 + \kd_2) \tau_n/2$ and $\alpha''_\pm = (\kd_1 + \kd_2) \tau_n/2 \pm (\ks_1 + \ks_2) \tau_n/2$, and the offset $(\ks_1+\ks_2)\tau_n/2$ in the measurement angle $\theta_n$, do not affect the final results for the decoherence. They did not appear in the special case because we deliberately chose $\kd_1 = -\kd_2 = \kd$ and $\ks_1 = -\ks_2 = \ks$ to simplify the expressions. This means that we can reuse some of the results in \cite{song2023optimized,tonekaboni2023greedy,liu2026imperfection}, by simply replacing $\kd$ with $(\kd_1 - \kd_2)/2$ and $\ks$ with $(\ks_1 - \ks_2)/2$, especially when the global phase factors do not contribute.

\subsection{Coherence vector for Map-based formalism}\label{sec:cohvector}
Given the structure of the map-based formalism to compute the DQ's coherence in Eq.~\eqref{eq:cohenoconH} and Eq.~\eqref{eq:maxcoh2}, we can define a coherence vector at any time $t_n$ to keep track of the state of knowledge of the unknown noise. For example, if there were $n$ measurements, then the coherence vector after the $n$th measurement is given by
\begin{equation}\label{eq:cohA}
    \ul{A}_n:={\bf F}(\mu_n,y_n) \cdots {\bf F}(\mu_2,y_2){\bf F}(\mu_1,y_1)\ul{P}_0,
\end{equation}
which means that we can write a recursive relation for the coherence vector as
\begin{equation}\label{eq:cohAn}
    \ul{A}_n={\bf F}(\mu_n,y_n)\ul{A}_{n-1}.
\end{equation}
Thus, the coherence in Eq.~\eqref{eq:maxcoh2} after $N$ measurements can be simplified to
\begin{equation} \label{eq:maxcohFmax}
    \cohc=\sum_{Y}\left|\ul{I}^{\top}\ul{A}_N\right|.
\end{equation}
%With the mapping matrix ${\bf F}(\mu_n,y_n)$, we can calculate the coherence of the DQ under a given measurement strategy ${\theta_n,\tau_n}$ and measurement outcome $y_n$. 
The resulting expression is similar to that for the RTP case in \cite{song2023optimized,tonekaboni2023greedy,liu2026imperfection}, with the only difference being the dimensions of ${\bf F}(\mu_n,y_n)$ and $\ul{A}_n$. 
%The next step is to determine the optimal measurement strategy that minimizes the decoherence of the DQ.
In this section, we analyze the coherence vector $\ul{A}_n$ in more detail. This vector encodes the relevant information about the environmental noise and will be used for the design of the measurement strategy in the subsequent section.

By comparing the coherence in Eq.~\eqref{eq:maxcoh} and Eq.~\eqref{eq:maxcohFmax}, if we define a complex conditional coherence as
\begin{equation}\label{eq:comAn}
    {\cal A}|_{\bullet}:=\left\langle e^{i\ul{\kd}^{\top} \ul{X}_n} \right\rangle_{\ul{X}_n|\bullet},
\end{equation}
then we can see a relationship between the probability functions and the coherence vector as
\begin{equation}\label{eq:sumAn}
    \wp(Y_n){\cal A}|_{Y_n}=\ul{I}^{\top}\ul{A}_n=\sum_{m_n}\wp(Y_n,m_n){\cal A}|_{Y_n,m_n}.
\end{equation}
Here we have defined the measurement results up to time $t_n$ as $Y_n :=\{ y_1,\cdots,y_n \}$, and the summation in the last equation is  over the noise levels $m_n\in \{1,2, \cdots, L\}$. With this, we find that the $L$ elements of $\ul{A}_n$ at time $t_n$ can be interpreted as
\begin{equation}\label{eq:cohAnele}
    \ul{A}_n=\left(\begin{matrix}
        A_n^{(1)}\\
        A_n^{(2)}\\
        \vdots\\
        A_n^{(L)}
    \end{matrix}\right)=\left(\begin{matrix}
        \wp(Y_n,m_n=1){\cal A}|_{Y_n,m_n=1}\\
        \wp(Y_n,m_n=2){\cal A}|_{Y_n,m_n=2}\\
        \vdots\\
        \wp(Y_n,m_n=L){\cal A}|_{Y_n,m_n=L}
    \end{matrix}\right).
\end{equation}
That is, the coherence vector $\ul{A}_n$ contains the probability of the measurement results $Y_n$ and $m_n$ and their associated complex conditional coherence.

The vector $\ul{A}_n$ does not give exactly the probability of the measurement results, $\wp(Y_n)$. However, if we take the limit of $\kappa\to 0$, this gives $\lim_{\kappa\rightarrow 0} {\cal A}_{|Y_n,m_n} = 1$ and $\lim_{\kappa\rightarrow 0} {\cal A}_{|Y_n}=1$, transforming Eq.~\eqref{eq:sumAn} to the probability distribution of $Y_n$,
\begin{equation}\label{eq:probYn}
    \wp(Y_n)=\lim_{\kappa\rightarrow 0}\ul{I}^{\top}\ul{{A}}_n \equiv \ul{I}^{\top}\ul{\check{A}}_n 
\end{equation}
Here we have also defined a probability vector,
\begin{equation}\label{eq:Ancheck}
    \ul{\check{A}}_n=\lim_{\kappa\rightarrow 0}\ul{A}_n=\left(\begin{matrix}
        \check A_n^{(1)}\\
        \check A_n^{(2)}\\
        \vdots\\
        \check A_n^{(L)}
    \end{matrix}\right)=\left(\begin{matrix}
        \wp(Y_n,m_n=1)\\
        \wp(Y_n,m_n=2)\\
        \vdots\\
        \wp(Y_n,m_n=L)
    \end{matrix}\right),
\end{equation}
which gives the joint probability of $Y_n$ and $m_n$ being in different levels $i$'s. In analogy to Eq.~\eqref{eq:cohAn}, the probability vector can also be updated after every SQ measurement via:
\begin{equation}\label{eq:Acheckupdate}
    \ul{\check{A}}_n=\check{\bf F}(\mu_n,y_n)\ul{\check{A}}_{n-1},
\end{equation}
if we defined a probability map with the same limit $\kappa\rightarrow 0$,
\begin{equation}\label{eq:Fcheck}
    \check{\bf F}(\mu_n,y_n):=\lim_{\kappa\rightarrow 0}{\bf F}(\mu_n,y_n).
\end{equation}
It is interesting to note that this $\check{\bf F}$ matrix contains only real numbers and thus can map between two real-number vectors containing only probability functions. Also, for the initial condition, we find that the coherence vector and the probability vector are simply the probability state at the initial time,
\begin{equation}
   \ul{ \check{A}}_0=\ul{A}_0=\ul{P}_0=\ul{P}_{\rm ss}.
\end{equation}
As per the final equality here, we are taking it to begin at steady-state from now on. %Each  
We also note that the probability vector $\ul{\check{A}}_n$ in Eq.~\eqref{eq:Acheckupdate} provides the information required for selecting the next measurement angle and waiting time, as discussed in the following section.

%%%%%%%%%%%%%%%%

\section{Adaptive algorithm for multi-level noise with single SQ}\label{sec:heuristic_adaptive_protocol}
In this section, we show how the map-based formalism and the coherence vector $\ul{A}_n$ can be used as a tool to design an adaptive protocol for measurement and control to improve the DQ's coherence under the multi-level noise. That is, we would like to search for the best measurement settings, i.e., the measurement angles $\{ \theta_1, \theta_2,..., \theta_N\}$, the waiting times $\{\tau_1, \tau_2, ..., \tau_N\}$, and their relationships with the coherence vector $\ul{A}_n$, such that the coherence is maximized. Because we are attempting to use a single SQ to probe $L$ levels of noise, we cannot hope to find the optimal algorithm, unlike for the $L=2$ case~\cite{song2023optimized,tonekaboni2023greedy}.  Thus we choose to work with a heuristic algorithm, which will be adaptive at every time step, 
informed by the $L=2$ results of those references.
 We will also analytically derive an upper bound on the decoherence rate for this adaptive algorithm at the end of this section.

\subsection{Heuristic adaptive algorithm}\label{sec:HAA}

Let us first motivate and summarize our proposed algorithm. Since each SQ's measurement can give at most one bit of information, the plausibly best use of that one bit is to distinguish the ``most-likely" noise level that the noise $m_t$ could be in from the ``second-most-likely" one. Suppose we have just performed the $n$-th measurement and need to know the best angle and time for the next measurement. Our algorithm for the measurement at $t_{n+1}$ will consist of (1) identifying the most-likely and the second most-likely levels, denoted by $\mu$ and $\nu$, respectively, (2) choosing the best measurement angle $\theta_{n+1}$ and the best measurement time $\tau_{n+1}$ that can maximally distinguish the two hypothetical states of  the SQ's given that the noise $m_t$ were in $\mu$ and $\nu$ during the measurement period.

\emph{Identifying likely noise levels}: Since $L$ vector elements of $\ul{\check{A}}_n$ are proportional to the likelihood of $m_n$ to be in different $L$ levels $j \in \{1, 2,..., L\}$ during the period $(t_n, t_{n+1})$, 
%The optimal angle we can choose is based on the most-likely noise value during this pair of measurements, which can be determined from the maximum element in the current coherence vector $\ul{ \check{A}}_n$. Without loss of generality, 
then we identify the most-likely noise $\mu$ as corresponding to the largest element of the vector $\ul{ \check{A}}_n$, i.e.,
\begin{align}\label{eq:lmu}
     \mu = {\rm argmax}_{j} \check{A}_n^{(j)}
\end{align}
%Since projective measurement of a qubit can only give one bit of information, we 
For the second most-likely noise level $\nu$, we consider two possible options. The first, most obvious, option is the index of the second largest element of the current $\ul{ \check{A}}_n$, denoted by $\alpha$. The second option, denoted by $\beta$, is the noise level with the highest transition probability from the most-likely noise level $\mu$ in Eq.~\eqref{eq:lmu}. However, to determine which option to choose for $\nu$, we come up with rough estimates of ``likelihood" of the two options by the time we make the next measurement. The likelihood of the first option is, roughly, the value of the second most-likely element, ${\check{A}}_n^{(\alpha)}$. For the second option, we estimate its likelihood by multiplying the current likelihood of the most-likely element, ${\check{A}}_n^{(\mu)}$, by its transfer rate $\gamma_{\mu\beta}$ to the level $\beta$ and the waiting time to the next measurement, $\pi/|\ks_\mu - \ks_\beta|$. (This waiting time is the time it takes for the SQ's phases, if the noise were in levels $\mu$ and $\beta$, to become maximally distinguishable; see below.) To summarize the two options and the criteria, we write
\begin{equation}\label{eq:seccon}
\nu=\left\{\,\,
\begin{aligned}
    &\, \alpha = {\rm argmax}_{j\ne \mu}\, \check{A}_n^{(j)}, \, &\text{if}~~ {\check{A}}_n^{(\alpha)} \, \geq \,\frac{\pi \, \gamma_{\mu\beta}\,{\check{A}}_n^{(\mu)} }{|\ks_{\mu}-\ks_{\beta}|},\\
    &\, \beta  = {\rm argmax}_j\, \gamma_{\mu j} \check{A}_n^{(\mu)}\blk,
     &\text{if}~~{\check{A}}_n^{(\alpha)} \, <\, \frac{\pi \, \gamma_{\mu\beta}\,{\check{A}}_n^{(\mu)} }{|\ks_{\mu}-\ks_{\beta}|}.
\end{aligned}
\right.
\end{equation}
%\hmw{check second line. Seems there was a term missing.}

\begin{figure}
    \centering
    \includegraphics[width=0.85\linewidth]{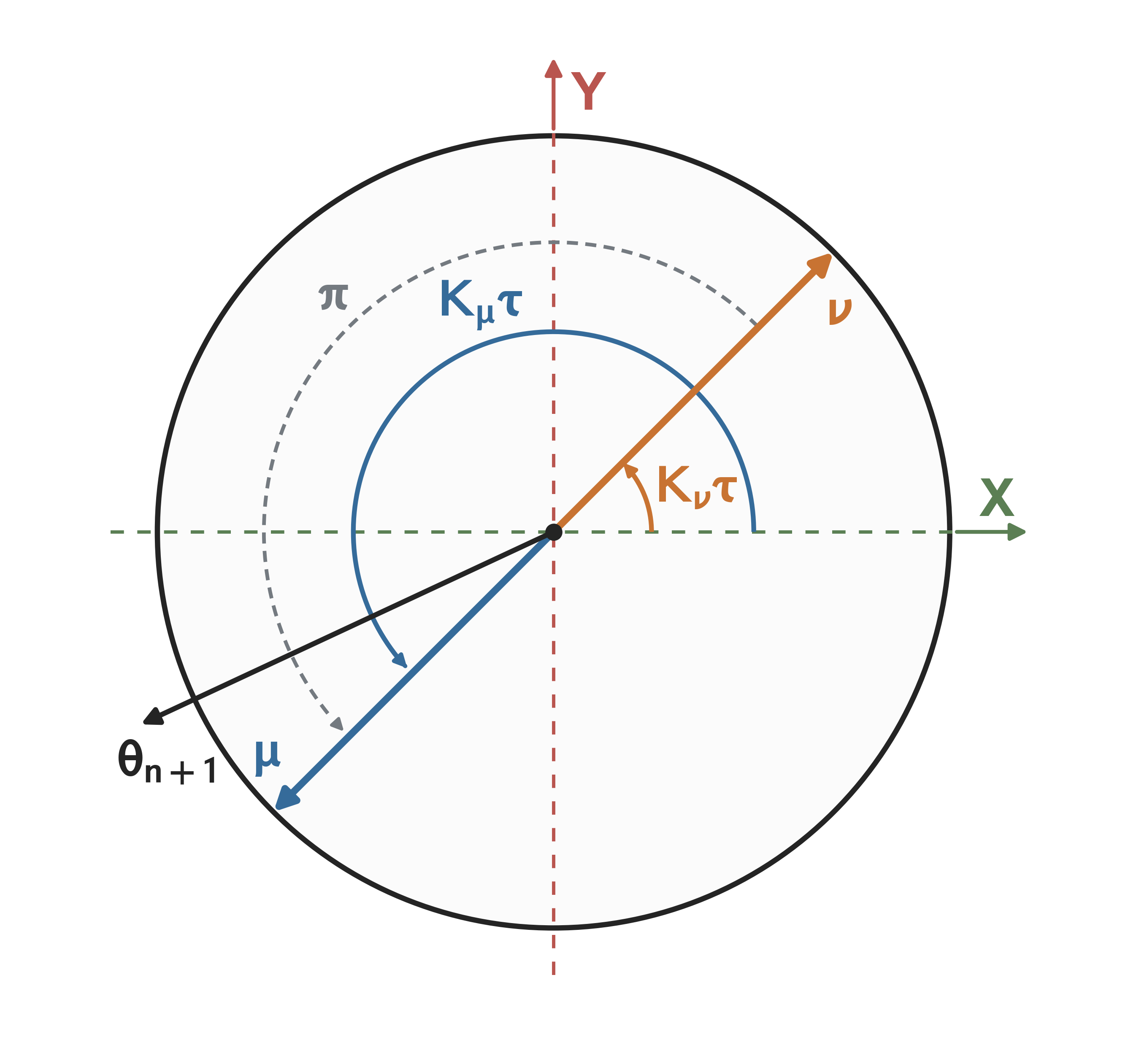}
    \caption{An illustration of the measurement angle and time that are used in the Heuristic Adaptive Algorithm. The orange and blue arrows describe the state vectors of the two plausible SQ's phases: $|\Phi\rangle^{\rm s} = |\ks_{\mu}\tau \rangle^{\rm s}$ and $|\Phi\rangle^{\rm s} = |\ks_{\nu}\tau\rangle^{\rm s}$, corresponding to the noise levels $\mu$ and $\nu$, respectively. The algorithm chooses $\tau_{n+1}$ based on the time when the two phases are at an angle $\pi$ apart as in Eq.~\eqref{eq:opttau}. The black arrow describes the measurement angle $\theta_{n+1}$, which is now shown to be at a random angle in the plot, but it should be equal to ${\rm max}\{ \ks_\mu \tau_{n+1}, \ks_\nu \tau_{n+1}\}$, i.e., either aligned with $|\ks_{\mu}\tau \rangle^{\rm s}$ or $|\ks_{\nu}\tau \rangle^{\rm s}$, following Eq.~\eqref{eq:theta}.}
    \label{fig:optimalangle}
\end{figure}

\emph{Choosing measurement settings}: 
Given the above analysis that the two most plausible noise values are $m_n = \mu$ and $m_n = \nu$, we can guess the two plausible SQ's phases after the period of waiting time $\tau_{n+1} = t_{n+1}-t_n$ to be $\ks_\mu \tau_{n+1}$ and $\ks_\nu \tau_{n+1}$, respectively.
%Projectively measuring on one single spectator qubit provides us only 1-bit of information. Therefore, we can use this 1-bit information to distinguish the``most-likely" noise out of the ``second most-likely" noise. We also denote the phase of the SQ corresponding to the ``most-likely" noise and the ``second most-likely" noise as the ``most-likely" phase and the ``second most-likely" phase, which are represented by $\ks_\mu \tau$ and $\ks_\nu\tau$. 
To maximally distinguish the two plausible phases with a single projective measurement on the SQ, we should wait until the waiting time $\tau_{n+1}$ is long enough such that the two phases are separated by $\pi$ (noting that $\ks_\mu \ne \ks_\nu$). See Fig.~\ref{fig:optimalangle} for the illustration. This implies that the waiting time should be
\begin{equation}\label{eq:opttau}
    \tau_{n+1}=\frac{\pi}{|\ks_\mu-\ks_\nu|},
\end{equation}
and the measurement angle should be able to distinguish the two phases, i.e., an angle that is orthogonal to the angle between the two phases,
\begin{equation}\label{eq:theta}
    \theta_{n+1}=\frac{\pi}{2}+\frac{(\ks_\mu +   \ks_\nu)\tau_{n+1}}{2},
\end{equation}
which can be simplified to $\theta_{n+1}=\max\{\ks_\mu\ddt_{n+1},\ks_\nu\ddt_{n+1}\}$ after substituting the waiting time in Eq.~\eqref{eq:opttau}. %We have summarized the algorithm in  pseudo-code in the Table~\ref{alg:heuristic}.

% 1: Initialise the first measurement time, $\tau_1$, and angle, $\theta_1$. Also initialise the coherence vector to start from $\ul{A}_0=\ul{\check{A}}_0=\ul{P}_{ss}$.\\
% 2: For each step $n$, get the index of the most-likely noise, $\mu$, from the maximal value in $\ul{\check{A}}_{n}$. Also get the index of the second most-likely noise, $\nu$, based on the condition in \eqref{eq:seccon}.\\
% 3: Set the next measurement time as $\tau_{n+1}=\frac{\pi}{|\ks_\mu-\ks_\nu|}$, and the measurement angle to be $\theta_{n+1}=\frac{\pi}{2}+\frac{\ks_\mu\ddt+\ks_\nu\ddt}{2}$.\\
% 4: Using the measurement result $y_{n+1}$ to update the coherence vector $\ul{A}_{n+1}$ and $\ul{\check{A}}_{n+1}$ with the $\ffun\left(\meas{=}\{\theta_{n+1},\ddt_{n+1}\},y_{n+1}\right)$ and $\check{\ffun}\left(\meas{=}\{\theta_{n+1},\ddt_{n+1}\},y_{n+1}\right)$, respectively.\\
% 5: Repeat for the step $n+1$.

\subsection{Upper bound for heuristic algorithm}
Before we show the numerical simulation results of the proposed Heuristic Adaptive Algorithm, in this section, we analytically derive the upper bound for the decoherence  of the DQ when the algorithm is applied. We start with the $L = 2$ special case, since we can compute the decoherence rate exactly for the asymptotic limit, where $T^{-1}, |\kd_{1,2}| \ll \gamma_{12,21} \ll |\ks_{1,2}|$. Following the calculation in Section~\ref{sec:nocontrol}, for $L=2$, and solving for the rate of change in no-control coherence when $t\rightarrow \infty$ to get $\cohnc(t) = \exp(-\Gamma^{\rm nc} t)$, we obtain the no-control decoherence rate given by
\begin{equation}\label{eq:ncdecoRTP}
\Gamma^{\rm nc}=\frac{(\kappa_1-\kappa_2)^2\check{\gamma}}{8\bar{\gamma}^2}.
\end{equation}
where $\check{\gamma}=(2\gamma_{12}\gamma_{21})/(\gamma_{12}+\gamma_{21})$ and $\bar{\gamma}=(\gamma_{12}+\gamma_{21})/2$, given the transition rates $\gamma_{12}, \gamma_{21}$ between the two noise levels. For the case with SQ's measurement and control, for $L=2$, we present here the results for a specific adaptive protocol in Refs.~\cite{tonekaboni2023greedy, song2023optimized}, in which $\theta_{n+1} = s_n \Theta + (\ks_1 + \ks_2)\tau/2$ and $\tau_{n+1} = 2\Theta/|\ks_1-\ks_2|$, where we have used the transformation in Eq.~\eqref{eq:oldnewH} and $s_n \in \{ +1, -1\}$ is chosen depending on some properties of $\ul{A}_n$. 
Following the calculation techniques in those works, we find that the decoherence rate is reduced to
%\hmw{Why do we do the optimal angle here? If we are only going to do the upper bound, we only need $\pi/2$.}
\begin{equation}\label{eq:cdecoRTP}
\Gamma=H_\Theta \frac{(\kd_1-\kd_2)^2 \check{\gamma}}{2(\ks_1-\ks_2)^2} = R(\Theta) \, \Gamma^{\rm nc},
\end{equation}
where $H_\Theta=3\Theta^2\csc^4(\Theta)\Theta-\left[2\Theta(\Theta-\cot(\Theta))+1 \right]\csc^2\Theta+\frac{1}{3}\Theta^2-1$. 
%noting that the rate becomes $\Gamma=H_\Theta  \kd^2 \check{\gamma}/2\ks^2$ for the special case $\kd_1 = - \kd_2 = \kd$ and $\ks_1 = -\ks_2 = \ks$ as presented in Refs.~\cite{tonekaboni2023greedy, song2023optimized}. 
The last equality of Eq.~\eqref{eq:cdecoRTP} expresses that the decoherence is suppressed from the no-control case by a reduction factor
\begin{equation}\label{eq:RRTP}
    R(\Theta) = \frac{4 H_\Theta\bar{\gamma}^2}{(\ks_1-\ks_2)^2}.
\end{equation}
The case we are interested in (which is very close to optimal for $L=2$~\cite{tonekaboni2023greedy,song2023optimized}) is when $\Theta=\pi/2$, which actually corresponds to the Heuristic adaptive algorithm in Eq.~\eqref{eq:theta}, where we have $\Theta = \theta_{n+1} - (\ks_1 + \ks_2)\tau/2 = \pi/2$ and the reduction factor becomes proportional to $H_{\Theta = \pi/2}$.
%where the coefficient $H_\Theta$ can be further minimized for the MOAAAR algorithm~\cite{tonekaboni2023greedy, song2023optimized}, with $\ks_1 = - \ks_2$, when choosing $\Theta=\Theta^\star\approx1.50055$, which gives $H_{\Theta^\star} \approx H^\star=1.254$.

For a general multi-level noise case, the analogue to the $L=2$  asymptotic regime is
\begin{equation}\label{eq:asymregime}
   {\rm max}_{j,k}\frac{|\kd_j-\kd_k|}{2}\,  \ll \,  \{ \gamma_{jk} \}\,  \ll \, {\rm min}_{j,k}\frac{|\ks_j-\ks_k|}{2} ,
\end{equation}
for any indices $j$ and $k$. While we cannot obtain an exact decoherence rate analytically when $L > 2$ under the Heuristic adaptive algorithm presented in Section~\ref{sec:HAA}, we can derive a plausible upper bound for it, in this asymptotic regime. The basic idea is to use the $L=2$ theory, but to multiply a worst-case no-control decoherence rate by a worst-case reduction factor $R$. 
%similar to the asymptotic regime in $L=2$ case. Using the fact that one projection measurement on the SQ provides one bit of information to only distinguish one pair of noise levels, we propose that a reasonable bound for decoherence comes from the dynamics of noise that switch between only two levels (out of $L$ levels) 

We begin with calculating the worst case no-control decoherence rate $\Gamma^{\rm nc}$. For $L$-level noise, the worst case is the decoherence rate the DQ would suffer if the RTP got ``stuck'' in the pair of levels that cause the greatest decoherence. That is, a plausible upper bound (ub) on $\Gamma^{\rm nc}$ can be obtained, based on  Eq.~\eqref{eq:ncdecoRTP}, by  maximizing over all pairs of noise $i,j \in \{1, 2, ..., L\}$:
\begin{equation}\label{eq:GamL}
\Gamma^{\rm nc}_{\rm ub}=\max_{i<j}\left\{\frac{(\kappa_i-\kappa_j)^2\check{\gamma}_{ij}}{8\bar{\gamma}_{ij}^2}\right\}.
\end{equation}
Here we have defined $\check{\gamma}_{ij}=(2\gamma_{ij}\gamma_{ji})/(\gamma_{ij}+\gamma_{ji})$ and $\bar{\gamma}_{ij}=(\gamma_{ij}+\gamma_{ji})/2$ analogously with the $L=2$ case. 

Next, we turn to finding the worst case reduction factor $R$. In this case, we also consider the $L=2$ expression, Eq.~\eqref{eq:RRTP}, and the worst two levels. To get an overall worse case, we do not take these worst-case levels to be the same as those in \erf{eq:GamL}, but rather choose them independently. For this reason we do not expect our upper bound to be tight at all; it is very conservative. Also, since, unlike \eqref{eq:ncdecoRTP}, Eq.~\eqref{eq:RRTP} is greater when the RTP transition rates are greater, we do not just ignore transitions out of the pair of levels, but rather keep the total transition rate out of a given level fixed, but force all the transitions to be into the worst (other) level. Again, this is gives a very conservative upper bound. 
 Thus, we obtain the upper bound on the  decoherence reduction factor for $L$-level noise, 
\begin{equation}\label{eq:RL}
R_{\rm ub}=\max_{i<j}\left\{H_{\pi/2} \times \left(\frac{\sum_{k\neq i}\gamma_{ik}+\sum_{k\neq j}\gamma_{jk} }{\ks_i-\ks_j}\right)^2\right\},
\end{equation}
%\bar{\gamma}_{i\rightarrow j}
maximizing over all pairs of noise $i,j$ as before.  
%We also replace the rate $\bar{\gamma}$ in Eq.~\eqref{eq:RRTP} with
%\beq
%\bar{\gamma}_{i\rightarrow j}=\frac{1}{2}\left(\sum_{k\neq i}\gamma_{ik}+\sum_{k\neq j}\gamma_{jk} \right),
%\eeq
%which is a simple average between the first term $\sum_{k\neq i}\gamma_{ik}$, the rate from $i$ to $j$, and the second term $\sum_{k\neq j}\gamma_{jk}$, the rate from $j$ to $i$, assuming all the noise went out of $i$ went to $j$, and vice versa.

Thus, we can conclude that a conservative upper bound for the decoherence rate for the multi-level noise is given by:
\begin{equation}\label{eq:analticbound}
\Gamma_{\rm ub}=R_{\rm ub}\, \Gamma^{\rm nc}_{\rm ub}, 
\end{equation}
where these factors are given be Eq.~\eqref{eq:GamL} and Eq.~\eqref{eq:RL}. It is important to note that the worst case decoherence rate scales like $\gamma \left(\Delta\kd/\Delta\ks\right)^2$, where $\gamma$ is the scale of the transition rates, $\Delta\kd$ the scale of the DQ sensitivity to different levels, and $\Delta\ks$ the scale of the SQ sensitivity to different levels. That is, the same scaling as \erf{eq:cdecoRTP} in the two-state RTP case. % \hmw{I checked. This was correct as it was, it was just referring to (48) where I should have referred to (47).}

\section{Numerical simulations for $L = 3$ multi-level noise}\label{sec:numericsim}
In this section, we consider an example of three-level noise $L = 3$, where the noise value $z(t)$ can be in any of the levels, $\{1, 2, 3\}$, with the transition rates $\gamma_{ij}$ between any levels $i$ and $j$. %The transition $J$ matrix is thus of the general form
We choose the following parameters for our numerical simulation:
\begin{equation}\label{eq:parameters}
\begin{aligned}
    \gamma_{12}=0.1, \gamma_{13}=0.2, \gamma_{23}=0.3,\\
    \gamma_{21}=0.2, \gamma_{31}=0.5, \gamma_{32}=0.3,\\
    \kd_1=0.004,\kd_2=0.001,\kd_3=0.005,\\
    \ks_1=40,\ks_2=100,\ks_3=180.\\
\end{aligned}
\end{equation}
%We start by identifying the probability vector
%\begin{equation}
%    \underline{P}_t=\begin{bmatrix}
%        \wp(m_t=1)\\
%        \wp(m_t=2)\\
%        \wp(m_t=3)
%    \end{bmatrix},
%\end{equation}
%for the probability of $m_t=i$ at time $t$ and the
%based on the asymptotic condition in \eqref{eq:asymregime}. 
From \erf{eq:Jmatrix}, these values lead to the transition matrix:
\begin{equation*}
    J=\begin{bmatrix}
        -0.3&0.2&0.5\\
        0.1&-0.5&0.3\\
        0.2&0.3&-0.8
    \end{bmatrix}.
\end{equation*}
We set the probability vector for $m_t=i$ at the initial time by the
asymptotic condition in \eqref{eq:asymregime}, then numerically simulate the Heuristic Algorithm. For completeness, we use three different angles and times for the first SQ measurement:\\\\
\emph{Initial conditions - Case 1}: 
\begin{align}
    \ddt_1=&\, \frac{\pi}{|\ks_1-\ks_2|}, \nonumber \\ \theta_1=&\, \frac{\pi}{2}+\frac{\ks_1\ddt_1+\ks_2\ddt_1}{2}=\ks_2\ddt_1 \nonumber
\end{align}
\emph{Initial conditions - Case 2}: \begin{align}
\ddt_1=&\,\frac{\pi}{|\ks_1-\ks_3|} \nonumber \\
\theta_1=&\,\frac{\pi}{2}+\frac{\ks_1\ddt_1+\ks_3\ddt_1}{2}=\ks_3\ddt_1 \nonumber
\end{align}
\emph{Initial conditions - Case 3}: \begin{align}
\ddt_1=&\, \frac{\pi}{|\ks_2-\ks_3|} \nonumber \\
\theta_1=&\, \frac{\pi}{2}+\frac{\ks_2\ddt_1+\ks_3\ddt_1}{2}=\ks_3\ddt_1 \nonumber
\end{align}
%Therefore, we can analyze the decoherence for the Heuristic algorithms for different values of measurement angles and waiting times.

\begin{figure}
    \centering
    \includegraphics[width=\linewidth]{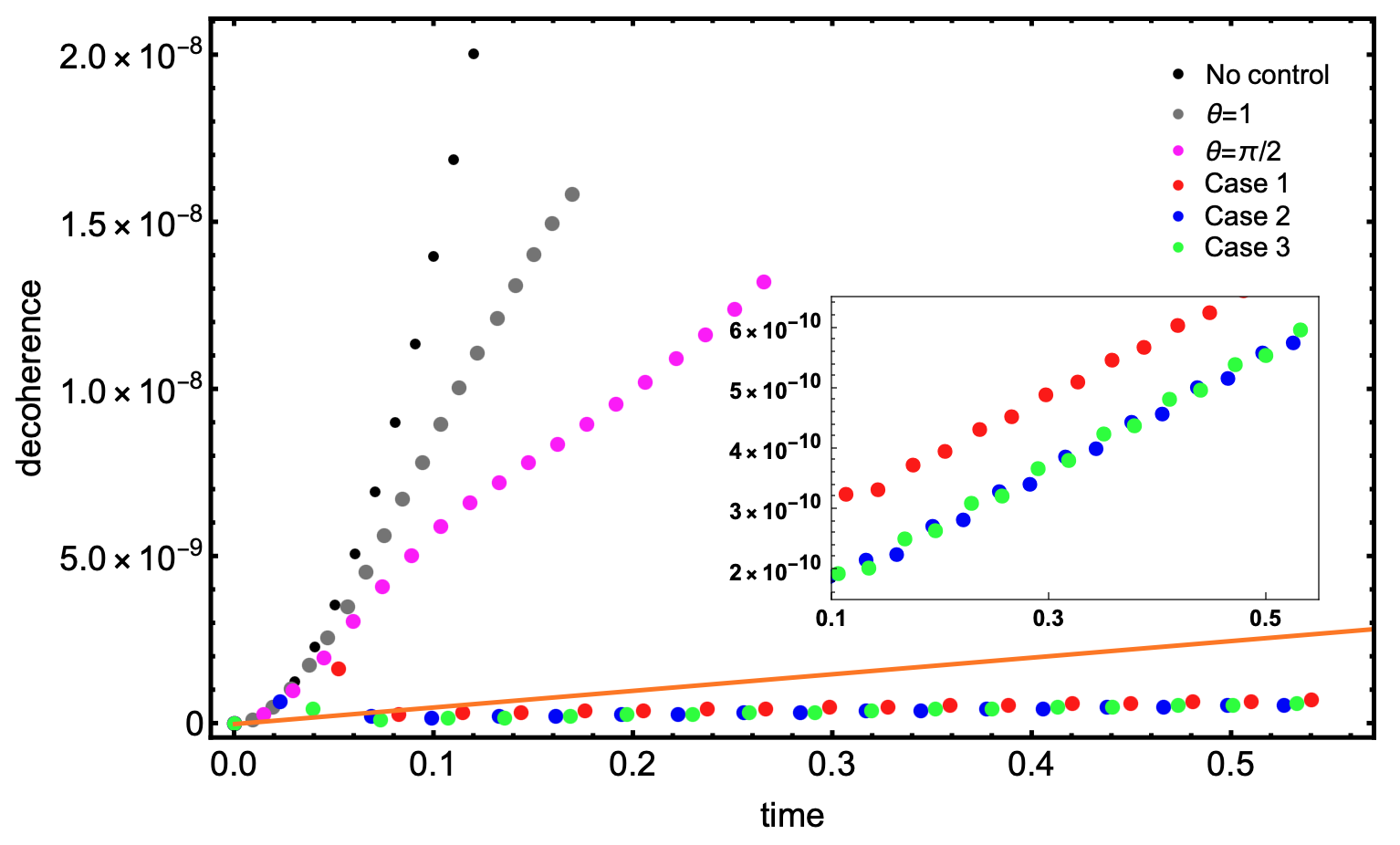}
    \caption{The decoherence as a function of time under different measurement strategies. The black dots correspond to the no-control case. The gray dots represent a fixed measurement angle and time: $\theta = 1$ and $\tau = \pi/\mathrm{Mean}(\ul{\ks})$. The magenta dots correspond to the fixed measurement angle and time: $\theta = \pi/2$ and $\tau = \pi/\mathrm{Mean}(\ul{\ks})$. The red, blue, and green curves show the results of the Heuristic Adaptive Algorithm with three initial measurement settings, i.e., case 1, case 2 and case 3. The orange solid line is the upper bound from \eqref{eq:analticbound}. The inset displays a zoomed-in view.}
    \label{fig:allresultswithzoomin}
\end{figure}

In Fig.~\ref{fig:allresultswithzoomin}, we show the numerical results for the decoherence as a function of time under
different measurement strategies. The decoherence is calculated as $1-\cohc$, with $\cohc$ as in \eqref{eq:maxcohFmax}. Comparing with the black dots for the no-control case, we can see clearly that the decoherence can be suppressed even for picked measurement angles and times picked with no optimization (see grey dots for $\theta = 1$ and pink dots for $\theta = \pi/2$ with $\tau = \pi/{\rm Mean}(\ul{K})$ where ${\rm Mean}(\ul{K}) \equiv (1/L)\sum_{i=1}^L \ks_i$). However, we obtain far greater suppression under our proposed Heuristic Adaptive Algorithm. Unsurprisingly, the performance of this Heuristic Adaptive Algorithm is not very sensitive to the choice of measurement in the first step, with data sets for case 1, case 2, and case 3 being very similar. We then calculate the decoherence rates of the three Heuristic cases by calculating their slopes, using data points from the $10$th measurement to the $18$th measurement, as they display a linear relation between decoherence and the time. We obtain three similar rates:  
$9.53 \times 10^{-10}$, $9.61 \times 10^{-10}$, and $9.55 \times 10^{-10}$ (in case number order). 

We also plot the analytical upper bound for multi-level random noise in Eq.~\eqref{eq:analticbound}, in solid orange in  Fig.~\ref{fig:allresultswithzoomin}. It can be seen that all the results using the Heuristic algorithm fit well below this upper bound, as expected. The decoherence for the non-optimal strategies (grey and pink dots), fall well above this bound, however. This shows that, while not being tight, it is still far from being a trivial upper bound.

\section{Conclusion}\label{sec:conclusion}
In this paper, we investigated a noise mitigation protocol using a sensitive spectator qubit (SQ) to probe a general multi-level noise process, in order to correct the dephasing in a nearby, less sensitive, data qubit (DQ). We modeled the noise as a random process that takes values in a finite set of $L$ levels, with arbitrary transition rates, and we allowed the rotation rate of each qubit to depend arbitrarily and independently on each level. 
To attack the problem, we extended the $L=2$ Bayesian map-based formalism of Refs.~\cite{song2023optimized,tonekaboni2023greedy} to arbitrary $L > 2$. This provides a maximally compact description of decoherence under the multi-level noise and reveals how the interplay between transition rates and coupling strengths determines the decay of coherence.
We analyzed the decoherence of the DQs without any control as well as how the decoherence can be reduced with measurements on SQs and a final DQ phase correction. 
%The Heuristic Adaptive Algorithm was proposed to select the measurement angle and waiting times, which greatly suppresses the decoherence rate of the DQs.
%The uncontrolled decoherence behavior was characterized, highlighting the contribution of multiple fluctuation channels and their associated timescales.

%In this work, we have investigated noise mitigation for a DQ subject to general multi-level fluctuator noise. The noise was modeled as a finite-state stochastic process with arbitrary transition rates, encompassing the conventional two-level random telegraph noise (RTP) model as a special case while capturing more realistic multi-state environments arising from multiple fluctuators. We first derived an analytical framework for the coherence dynamics of the DQ under such noise, based on the generator of the underlying Markov process. This formulation provides a unified description of decoherence for multi-level fluctuators and reveals how the interplay between transition rates and coupling strengths determines the decay of coherence. The uncontrolled decoherence behavior was characterized, highlighting the contribution of multiple fluctuation channels and their associated timescales.

%Building on this model, we analyzed a noise mitigation scheme using a SQ, where information about the noise is inferred through sequential measurements and used to perform phase corrections on the DQ. 
Based on intuition from the $L=2$ case, we propose we proposed a heuristic adaptive algorithm that dynamically selects the measurement time and angle based on the current estimate of the noise state. Numerical simulations demonstrate that the proposed approach can enormously suppress the decoherence rate of the DQ under multi-level noise in the parameter regime where the SQ is, in general, much more sensitive to that noise. In particular, the achieved performance has the same sort of scaling as that obtained in the two-level RTP case, despite the increased complexity of the noise process. Our findings extend the applicability of spectator-qubit-based noise mitigation beyond the commonly studied two-level setting and establish a general framework for analyzing and controlling decoherence under complex stochastic environments. %\hmw{beautiful.}

%\hmw{Areeya wants to explain the below more fully, and pose it more as future work on finding the best alogirithm in the degenerate case, and finding its performance. The main point I was trying to make is that there will still be some reduction $R<1$, just not $R$ very small.} 
Building on the findings of this work, several interesting questions remain for future investigation. One interesting question is the case when the SQ's sensitivities $\ul{\ks}_j$ are degenerate, while the DQ's sensitivities are non-degenerate. This means that the SQ cannot distinguish a pair or more RTP noise levels that would affect the DQ differently. %The exception to this is if $\ul{\ks}$ is degenerate for a pair of levels where $\ul{\kd}$ is non-degenerate. 
Then, based on Eq.~\eqref{eq:RL}, the reduction factor $R_{\rm ub}$  diverges, which is unphysical, but not unexpected because it is impossible to be in the regime of validity, \erf{eq:asymregime}. In this case, it is curious how much  degree of suppression (with a finite, but not so small $R_{\rm ub}$) would still be possible, from distinguishing the other noise levels outside of the degenerate ones. We expect that the algorithm adopted here would have to be modified, e.g., it would need to avoid accidentally choosing measurement angles and times such that $\ks_\mu = \ks_\nu$, that would result in diverging waiting time as in Eq.~\eqref{eq:opttau}.  

\begin{acknowledgments} \vspace{2ex}
This work was supported by the Australian Government via the Australia-US-MURI grant AUSMURI000002, by the Australian Research Council via the Centre of Excellence grant CE170100012. A.C.~also acknowledges the support of  the NSRF via the Program Management Unit for Human Resources and Institutional Development, Research and Innovation [grant number B39G680007].  
\end{acknowledgments}

%\bibliography{apssamp}% Produces the bibliography via BibTeX.
%apsrev4-2.bst 2019-01-14 (MD) hand-edited version of apsrev4-1.bst
%Control: key (0)
%Control: author (8) initials jnrlst
%Control: editor formatted (1) identically to author
%Control: production of article title (0) allowed
%Control: page (0) single
%Control: year (1) truncated
%Control: production of eprint (0) enabled
%

\begin{widetext}

\appendix
\section{Calculation of $\bf{H}$ matrix}\label{sec_app_Hmatrix}
In this section, we give the calculation details of the no-control mapping matrix $\bf{H}$.
We first calculate 
\begin{equation}\label{eq:probofX}
\begin{aligned}
    \wp(\ul{X},m_t|z_0)&=\sum_{m_1,m_2,\cdots,m_{M-1}}\wp(\ul{X}, m_t,m_{M-1},\cdots,m_1|m_0)=\sum_{m_1,m_2,\cdots,m_{M-1}}\wp(\ul{X},m_M,m_{M-1},\cdots,m_1|m_0)\\
    &=\sum_{m_1,m_2,\cdots,m_{M-1}}\delta(X_1-\sum_{k=1}^M\Delta t\delta_{1,m_k})\delta(X_2-\sum_{k=1}^M\Delta t\delta_{2,m_k})\cdots\delta(X_L-\sum_{k=1}^M\Delta t\delta_{L,m_k})\wp(m_M,m_{M-1},\cdots,m_1|m_0)\\
    &=\sum_{m_1,m_2,\cdots,m_{M-1}}\delta(X_1-\sum_{k=1}^M\Delta t\delta_{1,m_k})\delta(X_2-\sum_{k=1}^M\Delta t\delta_{2,m_k})\cdots\delta(X_L-\sum_{k=1}^M\Delta t\delta_{L,m_k})\wp(m_M|m_{M-1})\cdots\wp(m_1|m_0).
\end{aligned}
\end{equation}    
%\end{widetext}
Here, $\Delta t=t/M$ and $M$ is a very large number. In the second equation we defined $m_t=m_M$. Denote $H_{m_0}^{m_t}=\int \wp(\ul{X},m_t|m_0)e^{i\ul{\kd}^{\top}\ul{X}}d\ul{X}$, we can get 
%\begin{widetext}
\begin{equation}
\begin{aligned}
    H_{m_0}^{m_t}&=\int d\ul{X}e^{i\ul{\kd}^{\top}\ul{X}}\sum_{m_1,\cdots,m_{M-1}} \delta(X_1-\sum_{k=1}^M\Delta t\delta_{1,m_k})\delta(X_2-\sum_{k=1}^M\Delta t\delta_{2,m_k})\cdots\delta(X_L-\sum_{k=1}^M\Delta t\delta_{L,m_k}) \wp(m_M|m_{M-1}) \cdots \wp(m_1|m_0)\\
    &=\sum_{m_1,\cdots,m_{M-1}}e^{i\sum_{j=1}^L\kd_j(\sum_{k=1}^M\Delta t\delta_{j,m_k})} \wp(m_M|m_{M-1}), \wp(m_{M-1}|m_{M-2}) \cdots \wp(m_1|m_0)\\
    &=\sum_{m_1,\cdots,m_{M-1}}e^{i\Delta t \sum_{k=1}^M \sum_{j=1}^L\kd_j\delta_{j,m_k}} \wp(m_M|m_{M-1}) \wp(m_{M-1}|m_{M-2})\cdots\wp(m_1|m_0)\\
    &=\sum_{m_1,\cdots,m_{M-1}}e^{i\frac{\Delta t}{2}\sum_{j=1}^L\kd_j\delta_{j,m_M}}\left(e^{i\frac{\Delta t}{2}\sum_{j=1}^L\kd_j\delta_{j,m_M}} \wp(m_M|m_{M-1})e^{i\frac{\Delta t}{2}\sum_{j=1}^L\kd_j\delta_{j,m_{M-1}}}  \right)\cdots\\
    &\left(e^{i\frac{\Delta t}{2}\sum_{j=1}^L\kd_j\delta_{j,m_1}} \wp(m_1|m_{0})e^{i\frac{\Delta t}{2}\sum_{j=1}^L\kd_j\delta_{j,m_{0}}}\right)e^{-i\frac{\Delta t}{2}\sum_{j=1}^L\kd_j\delta_{j,m_0}}\\
    &=e^{i\frac{\Delta t}{2}\sum_{j=1}^L\kd_j\delta_{j,m_M}}\left( \sum_{m_1,\cdots,m_{M-1}}{\bf M}_{m_{M,M-1}},\cdots,{\bf M}_{m_1,m_0} \right)e^{-i\frac{\Delta t}{2}\sum_{j=1}^L\kd_j\delta_{j,m_0}}.
\end{aligned}
\end{equation}    
%\end{widetext}
Here in the last row of $H_{m_0}^{m_t}$, we have defined ${\bf M}_{m,m^\prime}=e^{i\frac{\Delta t}{2}\sum_{j=1}^L \kd_j\delta_{j,m}}\wp(m|m^\prime)e^{i\frac{\Delta t}{2}\sum_{j=1}^L \kd_j\delta_{j,m^\prime}}$. This ${\bf M}_{m,m^\prime}$ can be thought of as an element of a $L\times L$ matrix, where $m,m^\prime \in \{1,2,\cdots,L\}$. With the results of \eqref{eq:Ptsolution} and defining $T_{ij}=\wp(m=i|m^\prime=j)$, we obtain the following $L\times L$ matrix $\bf{M}$:
%\begin{widetext}
\begin{equation}\label{eq:bfMapp}
    \bf{M}=\begin{bmatrix}
        e^{i\kd_1\Delta t/2}&0&\cdots&0\\
        0&e^{i\kd_2\Delta t/2}&\cdots&0\\
        \vdots&\vdots&\ddots&\vdots\\
        0&0&\cdots&e^{i\kd_L\Delta t/2}
    \end{bmatrix}\begin{bmatrix}
        T_{11}&T_{12}&\cdots&T_{1L}\\
        T_{21}&T_{22}&\cdots& T_{2L}\\
        \vdots&\vdots&\ddots&\vdots\\
        T_{L1}&T_{L2}&\cdots&T_{LL}
    \end{bmatrix}\begin{bmatrix}
        e^{i\kd_1\Delta t/2}&0&\cdots&0\\
        0&e^{i\kd_2\Delta t/2}&\cdots&0\\
        \vdots&\vdots&\ddots&\vdots\\
        0&0&\cdots&e^{i\kd_L\Delta t/2}
    \end{bmatrix}.
\end{equation}    
%\end{widetext}

Using the above matrix, we can write $H_{m_0}^{m_t}=e^{i\frac{\Delta t}{2}\sum_{j=1}^L\kd_j\delta_{j,m_M}}({\bf M}^M)_{m_M,m_0}e^{-i\frac{\Delta t}{2}\sum_{j=1}^L\kd_j\delta_{j,m_0}}$. Therefore we have
%\begin{widetext}
\begin{equation}\label{eq:Hmatrixapp}
    {\bf H}=\lim_{M\rightarrow\infty}\begin{bmatrix}
        e^{i\kd_1\Delta t/2}&0&\cdots& 0\\
        0&e^{i\kd_2\Delta t/2}&\cdots&0\\
        0&0&\cdots&e^{i\kd_L\Delta t/2}
    \end{bmatrix}{\bf M}^M
    \begin{bmatrix}
        e^{-i\kd_1\Delta t/2}&0&\cdots& 0\\
        0&e^{-i\kd_2\Delta t/2}&\cdots&0\\
        0&0&\cdots&e^{-i\kd_L\Delta t/2}
    \end{bmatrix}.    
\end{equation}

\end{widetext}

\end{document}